# New Results for Radiative Proton Capture on $^{16}$O at Astrophysical Energies

## S B Dubovichenko* and A V Dzhazairov-Kakhramanov*

Fesenkov Astrophysical Institute "NCSRT" NSA RK,

050020, Observatory 23, Kamenskoe plato, Almaty, Kazakhstan

**Abstract :** The possibility of description of the experimental data for the astrophysical *S*-factor of the proton radiative capture on $^{16}$O to the ground and first excited states of $^{17}$F was considered in the frame of the modified potential cluster model with forbidden states and classification of the states according to Young tableaux. It was shown that on the basis of the *E*1 transitions from the *P* states of p$^{16}$O scattering to the bound states of $^{17}$F in the p$^{16}$O channel, it generally succeed to explain the value of measured astrophysical *S*-factors and reaction rates at astrophysical energies. The total cross sections, astrophysical *S*-factor from 10 keV and reaction rate at temperatures from 0.01 to 5 T$_9$.

## 1. Introduction

Previously on the basis of the modified potential cluster model (MPCM) [1–7] the possibility of description the coulomb form-factors of lithium nuclei was shown. This model uses intercluster potentials obtained on the basis of the elastic scattering phase shifts [8], taking into account the presence of the forbidden states (FSs) in wave functions (WFs) of system [9,10]. This approach was used by us still in the 80s in work [11], where FSs follow from the classification of cluster states according to Young tableaux [12]. Furthermore, in [13,14] in the MPCM with tensor forces the possibility of the correct reproduction practically of all characteristics of $^6$Li was demonstrated, including its quadrupole momentum. Subsequently, in [15–18] on the basis of the MPCM we have presented the possibility of description of the astrophysical *S*-factors or the total cross sections of almost 30 radiative capture processes, namely, *p*$^2$H, *n*$^2$H, *p*$^3$H, *p*$^6$Li, *n*$^6$Li, *p*$^7$Li, *n*$^7$Li, *p*$^9$Be, *n*$^9$Be, *p*$^{10}$B, *n*$^{10}$B, *p*$^{11}$B, *n*$^{11}$B, *p*$^{12}$C, *n*$^{12}$C, *p*$^{13}$C, *n*$^{13}$C, *p*$^{14}$C, *n*$^{14}$C, *n*$^{14}$N, *p*$^{15}$N, *n*$^{15}$N, *n*$^{16}$O and $^2$H$^4$He, $^3$He$^4$He, $^3$H$^4$He, $^4$He$^{12}$C systems at thermal and astrophysical energies (see also [2–7]).

Continuing study of thermonuclear reactions in the frame of the modified potential cluster model with forbidden states [15–18] let us consider the $^{16}$O(*p*,γ)$^{17}$F process, which takes part of the CNO cycle [15,16] and has additional interest, since it is the reaction at the last nucleus of 1*p*-shell with the forming of $^{17}$F that get out its limit. As we assume in [15–18], the bound state (BS) of $^{17}$F is caused by the cluster channel of the initial particles, which take part in the reaction.

Many stars, including the Sun, will eventually pass through an evolutionary phase that is referred to as the asymptotic giant branch [19]. This phase involves hydrogen and a helium shell that burn alternately surrounding an inactive stellar core. The $^{16}$O(p,γ)$^{17}$F reaction rate sensitively influences the $^{17}$O/$^{16}$O isotopic ratio predicted by models of massive (≥4$M_\odot$) Asymptotic Giant Branch (AGB) stars, where proton captures occur at the base of the convective envelope (hot bottom burning). A fine-tuning of the $^{16}$O(p,γ)$^{17}$F reaction rate may account for the measured anomalous $^{17}$O/$^{16}$O abundance ratio in small grains which are formed by the condensation of the material ejected from the surface of AGB stars via strong stellar winds [20].

## 2. Results and Discussion

### 2.1. Theoretical Calculation

---

* Corresponding author, E-mail: albert-j@yandex.ru

Continuing study the total cross sections of the radiative proton capture on $^{16}$O, let us firstly consider classification of orbital states for the p$^{16}$O system according to Young tableaux. The orbital Young tableau {4444} [12] corresponds to the ground bound state of $^{16}$O, therefore we have {1} × {4444} → {5444} + {44441} [21] for the p$^{16}$O system. The first of the obtained tableau is compatible with the orbital momentum $L = 0$ and is forbidden, because it could not be five nucleons in the $s$-shell, and the second tableau is allowed state (AS) and compatible with the orbital momentum $L = 1$ [22]. Thereby, in the potential of the $^2S_{1/2}$ wave, which corresponds to the first excited state (FES) of $^{17}$F at 0.4953 MeV with $J^\pi = 1/2^+$ relative to the ground state (GS) or -0.1052 MeV relative to the threshold of the p$^{16}$O channel and scattering states of these particles, there is the forbidden bound state. The $^2P$ scattering waves do not have the bound FSs, and the allowed state {44441} can be located in continuous spectrum. The ground state of $^{17}$F with $J^\pi, T = 5/2^+, 1/2$ in the p$^{16}$O channel is the $^2D_{5/2}$ wave at the energy -0.6005 MeV [23] relative to the threshold of the p$^{16}$O channel and also does not has forbidden bound states.

On the basis of data of $^{17}$F spectrum [23] one can consider that the $E$1 transition is possible from $^2P$ scattering waves with potential without FS to the $^2S_{1/2}$ FES of $^{17}$F with the bound FS.

1. $\begin{array}{l} ^2P_{1/2} \to\, ^2S_{1/2} \\ ^2P_{3/2} \to\, ^2S_{1/2} \end{array}$ .

Let us consider the $E$1 transition from the $^2P_{3/2}$ scattering wave with potential without FS for radiative capture to the $^2D_{5/2}$ GS without FS.

2. $^2P_{3/2} \to\, ^2D_{5/2}$ .

The GS and FES potentials will be constructed so that to describe the channel binding energy and the asymptotic constant (AC) of $^{17}$F in the p$^{16}$O channel correctly.

Elastic scattering phase shifts of the considered particles usually are used in the frame of the MPCM for the construction of cluster interaction potentials or for nucleons with nuclei [8]. Evidently, one of the first measurements of the differential cross sections of the p$^{16}$O elastic scattering with carrying out the phase shift analysis at energies of 2.0–7.6 MeV were done in [24]. This analysis used results [25] and [26] and some unpublished results [24] in the energy range 2.0–4.26 and 4.25–7.6 MeV, respectively. Furthermore, in [27], the polarizations of the p$^{16}$O elastic scattering in the energy range 2.0–5.0 MeV were measured and new phase shift analysis was done, which does not take into account the resonance at 2.66 MeV.

Furthermore, in works [28] and [29] the detailed phase shift analysis of the p$^{16}$O elastic scattering was carried out at energies of 1.5–3.0 MeV and the presence of the narrow resonance was shown for the specified subsequently energy of protons equals 2.663(7) MeV and the width of 19(1) keV, which corresponds to the first superthreshold state at 3.104 MeV with $J^\pi = 1/2^-$ relative to the GS or 2.5035 MeV in the center-of-mass system (c.m.) above the threshold of the p$^{16}$O channel [23]. Subsequently, the processes of the elastic scattering were considered in many works (see, for example, [23] and [30,31]) in the energy range 1.0–3.5 MeV. Particularly, in [32,33] the energies from 0.5–0.6 MeV and to 2.0–2.5 MeV were considered, but the phase shift analysis was not carried for them.

Since, we will further consider the radiative capture in the energy range up to 2.5 MeV, so the results of the listed works are quite enough for carrying out of the detailed phase shift analysis at energy, starting from 0.5 MeV. Furthermore, the potentials of the p$^{16}$O interaction will be constructed according to this phase shifts up to 2.5 MeV, i.e., without taking into account the first resonance with $J^\pi = 1/2^-$ at 2.663 MeV in laboratory system (l.s.) [23]. Thereto, the phase shift analysis of the available data was carried out at the energy range from 0.5 to 2.5 MeV [34]. Meanwhile, it was considered that in this energy range all $P$ and $D$ scattering phase shifts equal or approximate to zero.

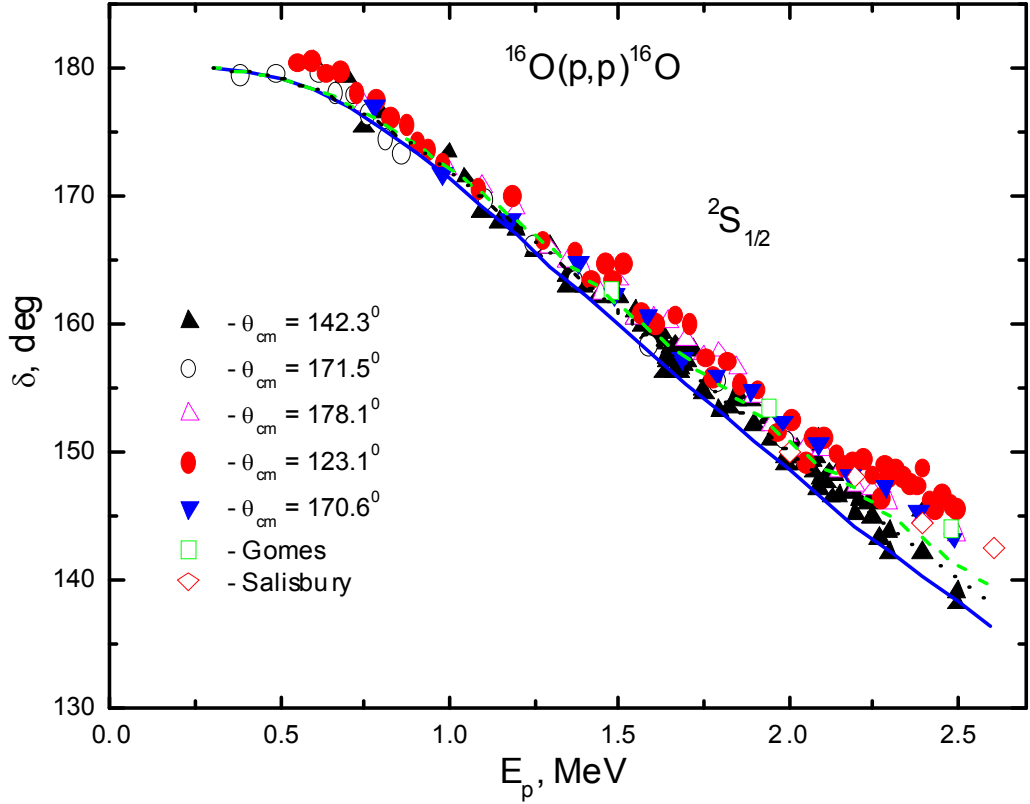

**Figure 1.** The p$^{16}$O scattering phase shifts, obtained in this work on the basis of the excitation functions at angles more than 120°. Experimental data are from [33,35–37]. Open rhombs and squares are results of the analysis at energies from 1.5 MeV to 2.5 MeV [24,28].

Therefore, only one scattering phase shift $^2S_{1/2}$ takes part in our analysis and, since in this wave in the considering energy range there are no resonances, it will have smoothly dropped shape, and results of such analysis are shown in Figure 1. Experimental data on excitation functions are from [33,35–37]. Our results are in a good agreement with obtained earlier phase shifts from [24,28] that are shown in Figure 1 by squares and rhombs at energies above 1.5 MeV.

2.2. Asymptotic Constants

Asymptotic normalization coefficient $A_{NC}$ (ANC) data are given, for example, in [38]. Here we will also use the known relation

$$A_{NC}^2 = S \times C^2, \qquad (1)$$

where $S$ is the spectroscopic factor; $C$ is the asymptotic constant in fm$^{-1/2}$, which join with the dimensionless AC [39] $C_W$, using by us further in the next way: $C = \sqrt{2k_0} C_W$, and the dimensionless value $C_W$ is defined by the expression [39]

$$\chi_L(r) = \sqrt{2k_0} C_W W_{-\eta L+1/2}(2k_0 r), \qquad (2)$$

where $\chi_L(r)$ is the numerical wave function of the bound state, obtaining from the solution of the radial Schrödinger equation and the integral over the wave function squared is normalized to unit; $W_{-\eta L+1/2}$ is the Whittaker function of the bound state that determines the asymptotic behavior of the wave function and that is the solution of the same equation without nuclear potential; $k_0$ is the wave number, stipulated by the binding channel energy $E$, $k_0 = \sqrt{\frac{2\mu}{\hbar^2}E}$; $\mu = \frac{m_1 m_2}{m_1 + m_2}$, $\eta$ is the coulomb parameter and $L$ is the orbital momentum of the bound state. Here, $\mu$ is the reduced mass of particles of the initial channel, and the constant $\hbar^2/m_0$ is equal to 41.4686 MeV fm$^2$, where $m_0$ – atomic mass unit (amu).

The radius of the GS of $^{16}$O equals 2.710(15) fm from [23] or 2.6991(52) fm from [40] was used in further calculations. For the GS and the FES of $^{17}$F the data for radii are absent [23,40], but they, apparently, should not differ a lot from the corresponding data for $^{16}$O. The charge proton radius and its mass radius are equal to 0.8775(51) fm [41]. The accurate values of nucleus and proton masses: $m(^{16}O)$ = 15.994915 amu [42] and $m(p)$ = 1.007276466812 amu [41] were used in all our calculations.

Furthermore, we will consider the radiative proton capture on $^{16}$O to the GS of $^{17}$F, which, as it was said, is the $^2D_{5/2}$ level and its potential has to describe the AC correctly. In order to extract this constant from the available experimental data, let us consider information about spectroscopic factors $S$ and asymptotic normalization coefficients $A_{NC}$. The obtained results for $A_{NC}$ are listed in Table 1, in addition, we succeeded in finding a lot of data of spectroscopic factors of $^{17}$F in the p$^{16}$O channel, and therefore we present their values in the form of separate Table 2.

Table 1. The data on $A_{NC}$ of $^{17}$F in the p$^{16}$O channel and astrophysical S-factors of the proton capture on $^{16}$O.

| Value of the $A_{NC}$ in fm$^{-1/2}$ for the GS | Value of the $A_{NC}$ in fm$^{-1/2}$ for the FES | $S(0)$ keV b for the GS | $S(0)$ keV b for the FES | $S(0)$ keV b, Total | References |
|---|---|---|---|---|---|
| 1.59 | 98.2 | --- | --- | 10.37 | [38] |
| 1.04(5) | 75.5(1.5) | 0.40(4) | 9.07(36) | 9.45(40) | [43] |
| 1.04(5) | 80.6(4.2) | 0.40(4) | 9.8(1.0) | 10.2(1.04) | [44] |
| 1.04(5) | --- | 0.317(25) | 8.552(43) | 8.869(44) | [36,45] |
| 1.10(1) | --- | --- | --- | --- | [35,45] |
| 1.19(2) | 81.0(9) | --- | --- | 7.1-8.2 | [45] |
| 1.13(1) | 82.3(3) | --- | --- | 7.1-8.2 | [45] |
| 0.97-1.09 | 86.4-91.1 | --- | --- | 10.2-11.0 | [46] |
| *0.97-1.59* | *74.0-98.2* | *0.29-0.44* | *8.7-10.8* | *7.1-11.06* | *Data Limit* |
| *1.28(31)* | *86.1(12.1)* | *0.37(7)* | *9.76(1.04)* | *9.08(1.98)* | *Average $\overline{A_{NC}}$ per interval* |

Table 2. Data on the spectroscopic factors $S$ of $^{17}$F in the p$^{16}$O channel of $^{17}$F

| $S$ for the GS | $S$ for the FES | References |
|---|---|---|
| 0.878 | 0.921 | [38] |
| 0.90(15) | 1.00(14) | Results of work [47] |
| 0.88 | 0.99 | Is given in [47] with ref. to other works |
| 0.94 | 0.83 | [23] |
| *0.88-1.05* | *0.83-1.14* | *Data Limit* |
| *0.97(9)* | *0.99(15)* | *Average $\overline{S}$ per interval* |

As it is seen from Table 2, the average values of the spectroscopic factors are close to unit, therefore, for ease, we will consider them equals 1. Furthermore, on the basis of eq. (1) for the GS we are finding that $\overline{A}_{NC}/\sqrt{S} = \overline{C}$ = 1.28 fm$^{-1/2}$, and since $\sqrt{2k}_0 = 0.57$, so dimensionless AC, determined as $\overline{C}_W = \overline{C}/\sqrt{2k_0}$, is equal to $\overline{C}_W$ = 2.25. However, the interval of the $A_{NC}$ values is so large that can lie in the limit 1.7–2.8. For the FES at $\sqrt{2k}_0 = 0.37$ we obtain the value $\overline{C}_W$ = 232.7 by analogy and the interval of the $C_W$ values taking into account $A_{NC}$ errors equal 200–265. These intervals can be expanded more if we will take into account the errors of the spectroscopic factors $S$ from Table 2.

2.3. Interaction Potentials

For carrying out the calculations of the radiative capture in the frame of the MPCM it is necessary to know potentials of the p$^{16}$O elastic scattering in the $^2P$ waves, and also interactions of the $^2D_{5/2}$ ground state and the $^2S_{1/2}$ first excited, but bound in the p$^{16}$O channel, state of $^{17}$F. There are experimental data for total cross sections of the radiative capture exactly for the transition to these BSs that were measured in [35,48]. Let us consider further the $E$1 transitions to the GS and FES and present the potentials of $^{17}$F in the p$^{16}$O channel too, since the phase shift value at energies from 0.5 to 2.5 MeV is known for the $^2S$ wave.

For example, for description of the obtained $^2S_{1/2}$ phase shift [34] as a result of our phase shift analysis it is possible to use the simple Gaussian potential with the point-like Coulomb part, FS and parameters [16]

$V_S$ = -75.02097 MeV,  $\gamma_S$ = 0.125 fm$^{-2}$, (3)

Energy dependence of the $^2S_{1/2}$ phase shift of potential from eq. (3) is shown in Figure 1 by the solid line, which starts from 180° [12] because of the presence of the FS. Such potential describes well the behavior of the $^2S$ scattering phase shift, obtained in our analysis, and quite agree with our previous extracting of the phase shifts [24,28]. At the same time, potential from eq. (3) allows to obtain the binding energy of the FES equals -0.105200 MeV at the accuracy of the finite-difference method (FDM) equals 10$^{-6}$ MeV [10], the charge radius of 3.10 fm, the mass radius of 2.93 fm of $^{17}$F, and the AC of the p$^{16}$O channel at the interval of 7–18 fm is equal to $C_W$= 215(2). This value is in a good agreement with the given above results of its extracting from $A_{NC}$ and spectroscopic factor $S$, laying at the region of the obtained above interval and is in a good agreement with results $\overline{C}_W$ = 204.0(4.1) of work [43].

The parameters for the $^2D_{5/2}$ GS potential of $^{17}$F in the p$^{16}$O channel without FSs were found

$V_D$ = -85.632465 MeV,  $\gamma_D$ = 0.12 fm$^{-2}$, (4)

which allow one to obtain the binding energy -0.600500 MeV at the FDM accuracy of 10$^{-6}$ MeV [10], the charge radius of 2.82 fm, the mass radius of 2.77 fm, and the AC at the interval of 6–25 fm is equal to $C_W$= 1.68(1) that is also at the lower limit of the given above interval for the AC. The scattering phase shift caused by this potential drops smoothly from zero and at 2.5 MeV equals the value approximately of -2°.

Since, there are no resonances of the negative parity in spectra of $^{17}$F at the energy lower than 2.5 MeV; we will consider that $^2P$ potentials should lead in this energy range practically to zero scattering phase shifts, and therefore they do not contain bound FSs, their depth can simple have a zero value.

2.4. The Astrophysical $S$-factor

Furthermore, the astrophysical $S$-factor of the proton capture on $^{16}$O were considered at the energy range up to 2.5 MeV. Results of our calculations for the $E$1 transitions to the GS with the potential from eq. (4) from $^2P$ scattering waves with potentials of zero depth in comparison with experimental data are shown in Figure 2 by the green solid line. The measurement results of the total cross sections in the range from 0.4 to 2.5 MeV [35] are shown by black squares, and circles are data from [48] at 0.5 – 2.5 MeV. Triangles show results of work [43], which lead to $S$-factor at zero energy of 0.4 keV·b.

The value of 0.41 keV·b at energy 100 keV was obtained in our calculations, which can be considered as zero energy, because the value of the $S$-factor at energies 90 – 150 keV has the value 0.410(1) keV·b. The calculated line at all energies completely lays within limits of the existent experimental errors of works [35,43,48]. The calculated $S$-factor at 30 keV rise up to 0.44 keV·b, therefore its values at the range 30 – 330 keV can be represented as 0.425(16) keV·b. As seen from Table 1, this value is in a good agreement with results of works [43,44]. The linear extrapolation of data [43] in the form

$$S(E) = 0.237E(\text{MeV}) + 0.40 \tag{5}$$

is shown in Figure 2 by the blue dashed line and leads to the results at zero energy of 0.40(1) keV·b. This line describes data [48] shown in Figure 2 by circles at energies up to 2.5 MeV quite well.

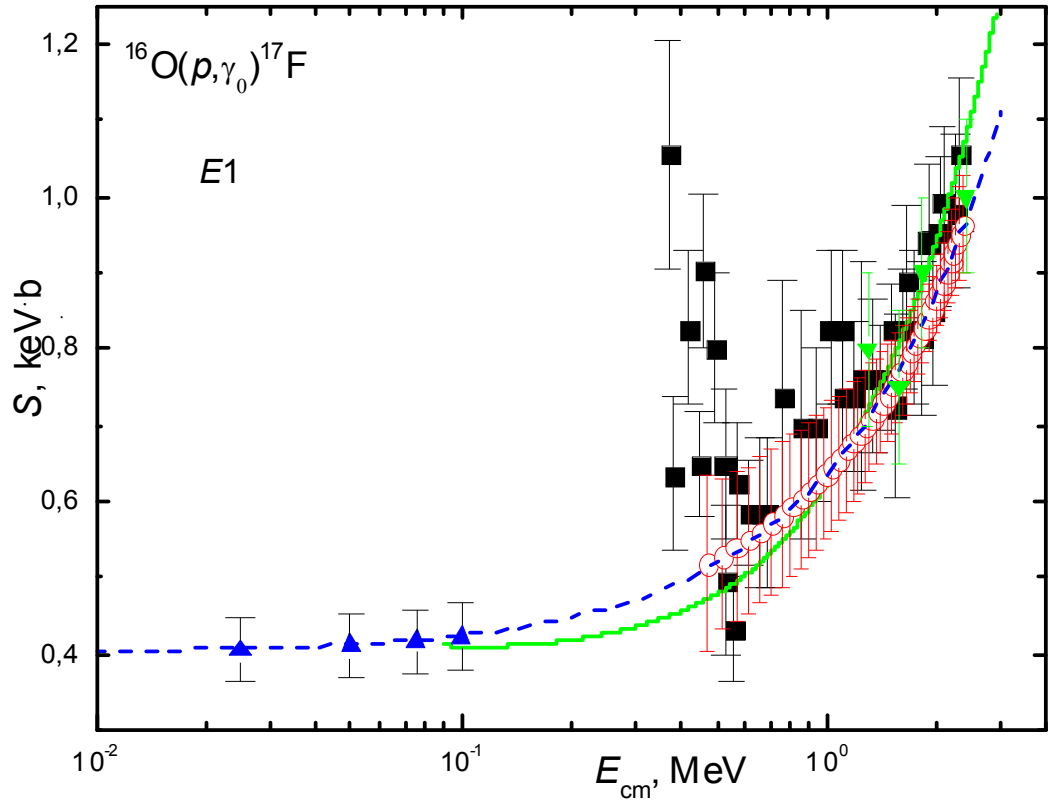

**Figure 2.** The astrophysical $S$-factor of the proton capture on $^{16}$O to the GS of $^{17}$F. Experimental data: squares – [35], red circles – [48], green triangles – [36], blue triangles – [43]. Lines are described in the text.

Thereby, we can see from Figure 2 that the carried out calculations of the $E$1 transition describe quite well the results of the experimental measurements of the astrophysical $S$-factor from [35,48] to the GS of $^{17}$F, laying in the range of experimental errors. Meanwhile, the potentials of the $^2P$ scattering waves, which do not contain FSs, are constructed on the basis of simple assumptions about coordination of such potentials with scattering phase shifts equal, in this case, to zero. The potential of the $^2D_{5/2}$ ground state of $^{17}$F in the p$^{16}$O channel was preliminarily coordinated with the basic characteristics of this nucleus, notably, with the binding energy and the AC in the p$^{16}$O channel.

Note that in our previous *S*-factor calculations [49] for the capture to the GS all matrix elements (ME) of transitions were calculated with the wave functions determined only up to 30 fm. Therefore, slightly other results were obtained for total cross sections at the capture to the GS. In this case ME of *S*-factors for the GS were calculated up to 200 fm, and for the FES up to 300 fm. Meanwhile, the scattering phase shift was obtained from the numerical WF at 30 fm, and at the large distances WF equates to its asymptotics. It allows one to obtain stable numerical results for *S*-factors at all considered energies.

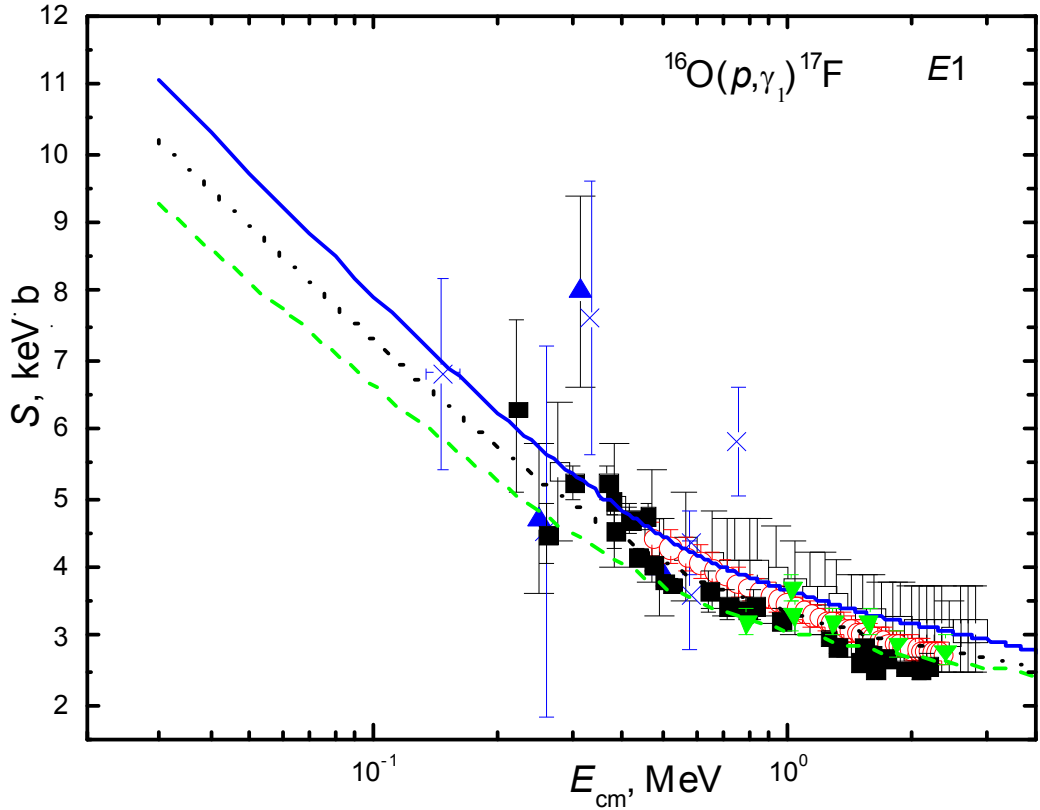

**Figure 3.** Total cross sections of the proton capture on $^{16}$O to the FES of $^{17}$F. Experimental data: open squares – [47], black squares – [35], blue crosses – [50], blue triangles – [51], red open circles – [48], green triangles – [36]. Lines are described in the text.

Furthermore, the *S*-factor for capture to the FES was shown in Figure 3 by the blue solid line. Potential (3) and zero variants for *P* waves were used. The *S*-factor has the value of 11.07 keV·b at energies 30 keV and at major energies agrees well with the results of experiments [47–51]. For example, in the Resonating Group Method (RGM) for zero energy was obtained 10–11 keV·b [46].

If to coordinate calculation results of the *S*-factor for capture to the FES with data of work [35], as it was done in [43], then they were shown in Figure 3 by the green dashed line, which has the value of 9.27 keV·b at 30 keV – the value of 9.07(36) keV·b was obtained in [43]. These potential parameters were used

$$V_S = -87.855325 \text{ MeV}, \quad \gamma_S = 0.15 \text{ fm}^{-2}, \tag{6}$$

Energy dependence of the $^2S_{1/2}$ phase shift is shown in Figure 1 by the green dashed line. The potential allows one to obtain the binding energy equals -0.105200 MeV with the FDM accuracy of $10^{-6}$ MeV [10], the charge radius of 3.05 fm, the mass radius of 2.90 fm of $^{17}$F, and AC of the p$^{16}$O channel at the interval of 5–20 fm is equal to $C_W$ = 197(2). This value is in the minimum limit of the obtained above interval.

It is possible to consider another variant of calculations, which results are shown in Figure 3 by the black dotted line. The *S*-factor at 30 keV has the value 10.19 keV·b, and these values were used

for potential parameters

$$V_S = -82.74659 \text{ MeV}, \quad \gamma_S = 0.14 \text{ fm}^{-2}, \qquad (7)$$

Energy dependence of the $^2S_{1/2}$ phase shift is shown in Figure 1 by the black dotted line. The potential allows one to obtain binding energy equaling -0.105200 MeV with the FDM accuracy of $10^{-6}$ MeV [10], the charge radius of 3.07 fm, the mass radius of 2.91 fm of $^{17}$F, and AC of the p$^{16}$O channel at the range of 6–18 fm is equal to $C_W$ = 204(2). These results slightly better agree with data [48] and RGM results for zero energy [46].

Reaction rates of the proton capture on $^{16}$O are given in Figure 4. Potentials (3) and (4) were used for calculations of these rates, and total cross sections were calculated at energies from 30 keV to 5 MeV, but without taking into account resonances above 2.5 MeV.

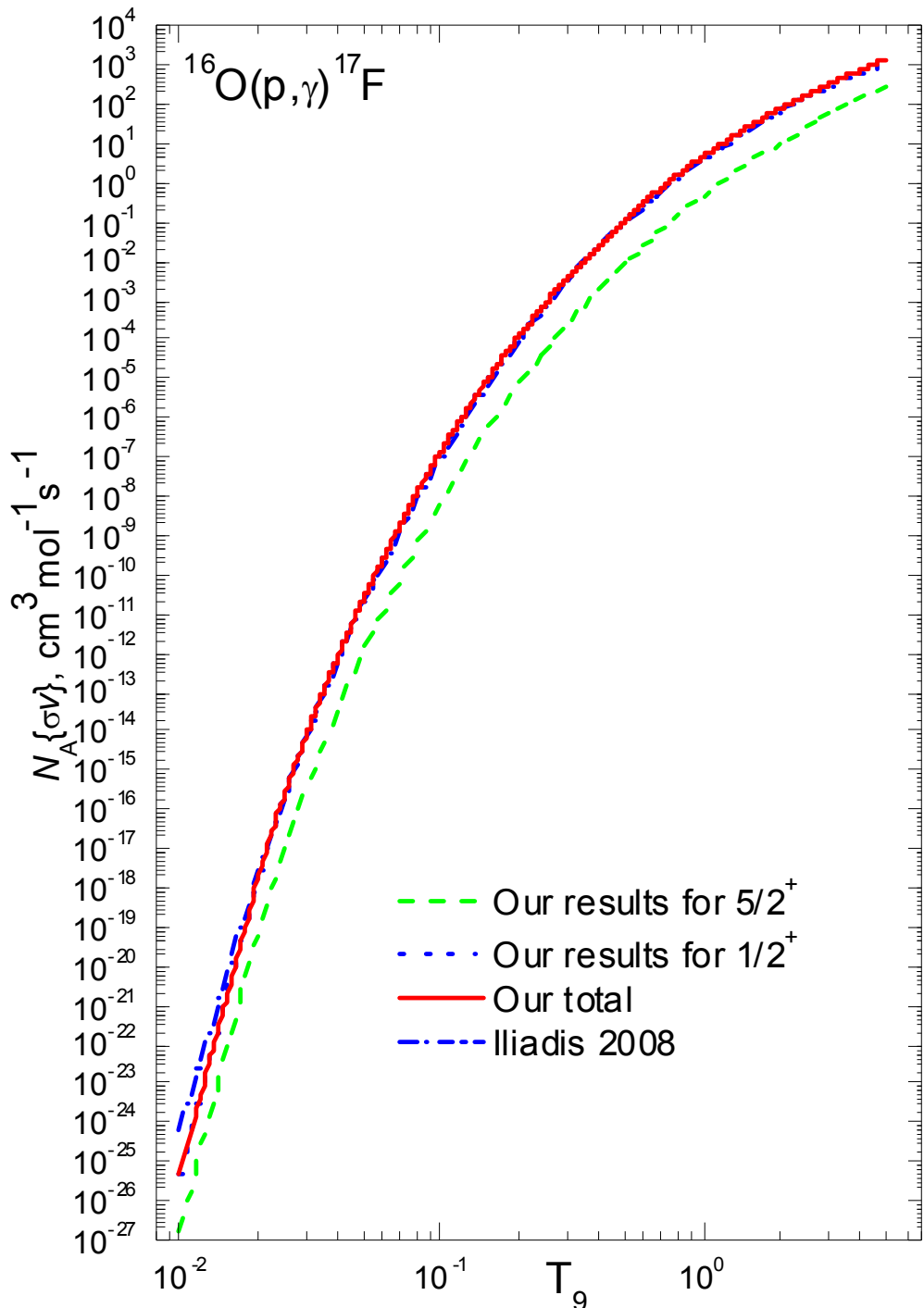

**Figure 4.** Reaction rate of the proton capture on $^{16}$O. The dotted-dashed line is in from [45].

The reaction rate is written in the usual form [52]

$$N_A \langle \sigma v \rangle = 3.7313 \cdot 10^4 \mu^{-1/2} T_9^{-3/2} \int_0^\infty \sigma(E) E \exp(-11.605 E / T_9) dE,$$

where $E$ is given in MeV, the cross section $\sigma(E)$ assigned in $\mu$b, $\mu$ is the reduced mass in aum, $T_9$ – temperature in $10^9$ K.

The calculation results of the reaction rate from [45], which practically coincide with our results, are shown in Figure 4 by the blue dotted-dashed line. Small difference exists only at temperatures lower 0.02–0.03 $T_9$.

### 3. Conclusions

Thereby, the considered above methods of constructions of interaction potentials of clusters allow one generally correctly reproduce experimental data for the astrophysical $S$-factor of the radiative capture at energies from 0.03 MeV to 2.5 MeV. It was obtained the potential of the $^2D_{5/2}$ scattering wave without FS with the phase shift approximate to zero, which is used for the GS and describes its basic characteristics. The potential for the $^2S_{1/2}$ wave with FS correctly reproduces the energy dependence of the scattering phase shift obtained in the result of our phase shift analysis and agrees with the fundamental characteristics of the FES.

### Acknowledgments

This work was performed under grant No. 0070/GF4 "Thermonuclear reactions in stars and controlled thermonuclear fusion" through Fesenkov Astrophysical Institute "NCSRT" ASA MDASI RK.